\documentclass[11pt]{article}
\usepackage{moriond,epsfig}

\def\Journal#1#2#3#4{{#1} {\bf #2}, #3 (#4)}

\def\PLB{{\em Phys. Lett.}  B}
\def\PRL{\em Phys. Rev. Lett.}
\def\PRC{{\em Phys. Rev.} C}

\def\be{\begin{equation}}
\def\ee{\end{equation}}
\def\bea{\begin{eqnarray}}
\def\eea{\end{eqnarray}}

\begin{document}
\vspace*{4cm}
\title{Systematics of Soft Particle Production at RHIC: Lessons from PHOBOS}

\author{G.\ S.\ F.\ Stephans$^4$ for the PHOBOS Collaboration\\
B.Alver$^4$, B.B.Back$^1$, M.D.Baker$^2$, M.Ballintijn$^4$, D.S.Barton$^2$, R.R.Betts$^6$,
A.A.Bickley$^7$, R.Bindel$^7$, W.Busza$^4$, A.Carroll$^2$, Z.Chai$^2$, V.Chetluru$^6$, 
M.P.Decowski$^4$, E.Garc\'{\i}a$^6$, T.Gburek$^3$, N.George$^2$, K.Gulbrandsen$^4$, 
J.Hamblen$^8$, I.Harnarine$^6$, C.Henderson$^4$, D.J.Hofman$^6$, R.S.Hollis$^6$, 
R.Ho\l y\'{n}ski$^3$, B.Holzman$^2$, A.Iordanova$^6$, E.Johnson$^8$, J.L.Kane$^4$, 
N.Khan$^8$, P.Kulinich$^4$, C.M.Kuo$^5$, W.Li$^4$, W.T.Lin$^5$, C.Loizides$^4$, 
S.Manly$^8$, A.C.Mignerey$^7$, R.Nouicer$^{2,6}$, A.Olszewski$^3$, R.Pak$^2$, C.Reed$^4$, 
E.Richardson$^7$, C.Roland$^4$, G.Roland$^4$, J.Sagerer$^6$, P.Sarin$^4$, I.Sedykh$^2$, 
C.E.Smith$^6$, M.A.Stankiewicz$^2$, P.Steinberg$^2$, A.Sukhanov$^2$, A.Szostak$^2$, 
M.B.Tonjes$^7$, A.Trzupek$^3$, C.Vale$^4$, G.J.van~Nieuwenhuizen$^4$, S.S.Vaurynovich$^4$, 
R.Verdier$^4$, G.I.Veres$^4$, P.Walters$^8$, E.Wenger$^4$, D.Willhelm$^2$, F.L.H.Wolfs$^8$, 
B.Wosiek$^3$, K.Wo\'{z}niak$^3$, S.Wyngaardt$^2$, B.Wys\l ouch$^4$\\
\small 
$^1$~Argonne National Laboratory, Argonne, IL 60439-4843, USA\\
$^2$~Brookhaven National Laboratory, Upton, NY 11973-5000, USA\\
$^3$~Institute of Nuclear Physics PAN, Krak\'{o}w, Poland\\
$^4$~Massachusetts Institute of Technology, Cambridge, MA 02139-4307, USA\\
$^5$~National Central University, Chung-Li, Taiwan\\
$^6$~University of Illinois at Chicago, Chicago, IL 60607-7059, USA\\
$^7$~University of Maryland, College Park, MD 20742, USA\\
$^8$~University of Rochester, Rochester, NY 14627, USA\\}

\maketitle
\abstracts{
The PHOBOS experiment has measured the properties of particle
production in heavy ion collisions between $\sqrt{s_{_{NN}}}$ of 20 and 200~GeV.
The dependencies of charged particle yield on energy, system size, and
both longitudinal and transverse momentum have been determined over close
to the full kinematic range.  Identified charged particles emitted near
mid-rapidity have been studied over about 2 orders of magnitude in
transverse momentum.  This broad data set was found to be characterized by
a small number of simple scalings which factorize to a surprising degree.
This study has recently been extended by the addition of new data for
Cu+Cu as well as new analyses of Au+Au data, including more peripheral
collisions.  In addition, the exploration of global properties has been
expanded with the use of new techniques, including two-particle
correlations, more sensitive searches for rare events, and more detailed
studies of particles emitted at very forward rapidity.  The characteristics
of particle production which are revealed by this extensive data set will
be described along with the implications for future data from the LHC.
}

\vspace{-8pt}
\section{Introduction} 

The PHOBOS experiment took data at RHIC starting with the first beam in June
of 2000 and continuing through Run 5 in the spring of 2005.  Data were taken
for a broad range of systems, namely p+p at two energies, d+Au at one energy,
Cu+Cu at three energies, and Au+Au at five energies. Results from PHOBOS and
the three other RHIC experiments have shown that heavy ion collisions at the
highest RHIC energies result in the formation of a new state of matter,
characterized by a high energy density and dominated by partonic degrees of
freedom.\cite{RHICWhitePaper}

One of the primary goals of the PHOBOS experimental program was the
characterization of the properties of particle production over a very broad
range in energy and system size, as well as over several orders of magnitude
in transverse momentum and all or a very large fraction of the pseudorapidity
distribution. While not necessarily evidence for, or a direct probe of, the exotic
partonic state, these observables set constraints on models of the formation
and subsequent hadronization of the novel medium.  Final state particle
distributions can also set limits on basic properties of the system such as
energy density and entropy.  In addition to contributing significantly to our
understanding of the systems formed at RHIC, this extensive data set has
revealed a number of surprising results.

\vspace{-8pt}
\section{Energy Dependence of Particle Production}

The first physics result from RHIC was the PHOBOS publication of the charged
particle pseudorapidity density, $dN/d\eta$, near mid-rapidity at nucleon
nucleon center of mass energies ($\sqrt{s_{_{NN}}}$) of 56 and
130~GeV.\cite{Phobos1} This early result had three immediate and profound
impacts on the field of relativistic heavy ions. First, the numerical value
invalidated the majority of the theoretical predictions in existence at the
time.  Second, the data lent support to concepts of parton saturation, which
if validated could describe the dominant physics process controlling the low-x
region at high energies, even in p+p collisions.  Finally, the fact that these
values were significantly lower than many of the theoretical predictions,
combined with the first PHOBOS $dN/d\eta$ data at $\sqrt{s_{_{NN}}}$=200~GeV
which was also on the low side of the revised theoretical
predictions,\cite{PhobosFirst200} suggested that tracking and other
measurements in heavy ion collisions at the LHC might not be as formidable as
originally thought.  This realization helped to spawn a significant expansion
in the planned LHC heavy ion program.

Later analysis revealed an intriguing similarity between the particle
multiplicities per pair of participating nucleons in nucleus-nucleus collisions when compared
to proton-(anti)proton interactions at twice the center of mass energy 
(i.\ e.\ (p+p)$\sqrt{s}$=2$\times$(A+A)$\sqrt{s_{_{NN}}}$).  Further, these multiplicities were
similar to those seen in e$^+$+e$^-$ at the same energy.\cite{Universality}
The comparison of p+p and e$^+$+e$^-$ was known previously and assumed to be
due to the fact that only about half of the center of mass energy in p+p was
available for particle production. The new comparison with Au+Au implies that
nucleus-nucleus collisions can convert a much larger fraction of the available
energy into particles.  Data from the LHC (p+p at 14~TeV and Pb+Pb at 5.5~TeV)
will reveal whether or not this correspondence extends to much higher
energies.

\vspace{-8pt}
\section {Pseudorapidity Dependence of Particle Production}

\begin{figure}
\begin{minipage}{7.0cm}
\begin{center}
\epsfig{figure=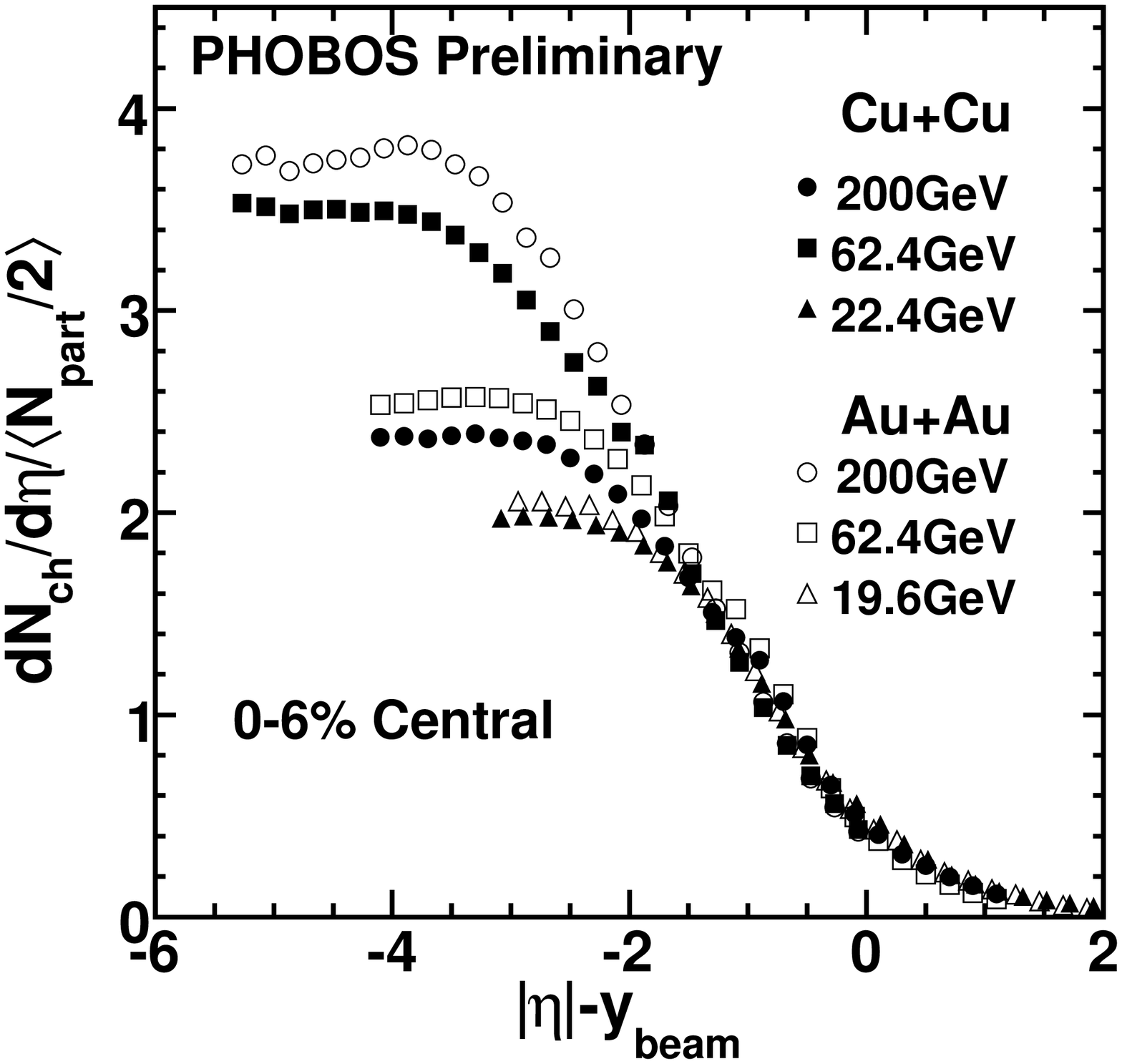,width=5.8cm}
\epsfig{figure=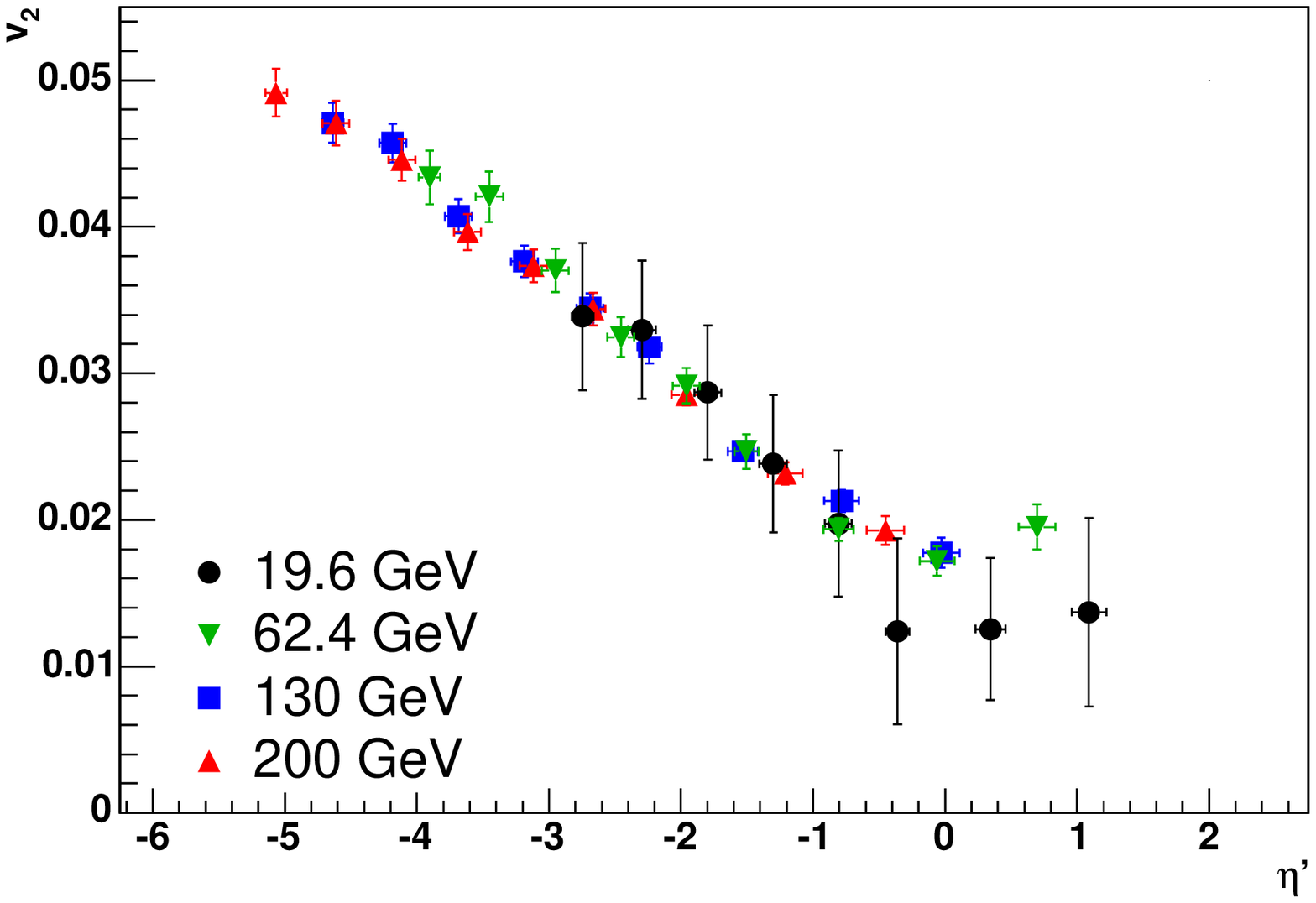,width=6.3cm}
\end{center}
\end{minipage}
\begin{minipage}{7.8cm}
\begin{center}
\epsfig{figure=s_cent_scaling_dNdeta.eps,width=7.8cm}
\end{center}
\end{minipage}
\caption{(Top left) Charged particle $dN/d\eta$ for various systems and energies
effectively viewed in the rest frame of one of the colliding nuclei. (Bottom 
left) A similar plot for elliptic flow. (Right) $dN/d\eta$
per participant pair versus centrality for Au+Au at four energies fit using a
product of separate functions of energy and centrality.}
\label{OneFig}
\end{figure}

The uniquely broad pseudorapidity coverage of the PHOBOS multiplicity detector
allowed measurement of all or almost all of the $dN/d\eta$ distribution, even 
at the highest RHIC energies.  In addition to producing
total multiplicity data with relatively small systematic errors, these results
made possible a detailed comparison of the shape of the distribution at
different center of mass energies.  When the $dN/d\eta$ distributions for
Au+Au ranging from $\sqrt{s_{_{NN}}}$=19.6 to 200~GeV were plotted as a
function of $\eta-y_{beam}$, thereby effectively viewing them in the rest
frame of one of the colliding nuclei, it was found that data from all energies
followed a common curve (see top left panel of 
Fig.~\ref{OneFig}).\cite{PhobosLimFrag1} \cite{PhobosMult64} Furthermore,
preliminary data for Cu+Cu over roughly the same range in energy reveal that
they follow exactly the same curve.\cite{PhobosPlenaryQM06} Thus, the ``limiting
fragmentation" or ``extended longitudinal'' scaling seen previously in small
systems was found to apply also in heavy ion collisions.  Assuming that this 
observation and the energy
dependence described in the previous section extend up to LHC energies, an
empirical prediction of the full $dN/d\eta$ shape can be made.\cite{BuszadNdeta}

A related, but much more surprising, result was found when a similar analysis
was applied to the pseudorapidity dependence of elliptic flow.  When the data
for v$_2$ from Au+Au were plotted in the effective rest frame of one of the
colliding nuclei, a pattern of ``extended longitudinal scaling" was again
revealed (see bottom left panel of Fig.~\ref{OneFig}.\cite{Phobosv2eta} 
This adds an additional intriguing element to the experimental results
for the pseudorapidity dependence of elliptic flow, data which
have presented a considerable challenge to existing theories (see for example
\cite{FlowTheory}).  Again, an
extrapolation to LHC energies can be done but the interpretation of the result
is unclear. The magnitude of v$_2$ at midrapidity in Au+Au at
$\sqrt{s_{_{NN}}}$=200~GeV is claimed to saturate the hydrodynamical
limit but, if the observed trend continues, the value at the LHC will be
significantly larger. 

\vspace{-8pt}
\section {Centrality Dependence of Particle Production}

By analyzing heavy ion collisions at varying impact parameter, the effect of
system size on particle production can be explored.  This variation does not
represent simply ``more of the same" since central collisions with small
impact parameters have a larger average number of collisions per participant
and therefore might differ more significantly from elementary p+p
interactions. A common claim is that ``harder" processes should scale with the
number of collisions while ``softer" process should scale with the number of
participants. If true, this belief, combined with the expectation that the
ratio of ``hard" over ``soft" processes should increase with collision energy,
implies that the centrality dependence must be energy dependent. In stark
contrast to this prediction, the centrality dependence is found to be
identical at all energies studied.\cite{PhobosTracklet} \cite{PhobosMult64} In
fact, it can be shown quantitatively that the data factorize by fitting them
with the product of separate functions of energy and the number of
participating nucleons (see right panel of 
Fig.~\ref{OneFig}).\cite{BirgerFit} Far from being a property solely of
bulk particle production at lower transverse momentum, this factorization was
found to extend up to p$_T$ of almost 4~GeV/c in Au+Au
data.\cite{PhobosSpectraCent} Again, Cu+Cu results are observed to follow the
same trend.\cite{PhobosSpectraCuCu} As with many of
the PHOBOS observations, it will be very interesting to follow this trend to
the LHC where ``hard" processes are expected to make a much more dominant
contribution to particle production.

\vspace{-8pt}
\section {Continuing PHOBOS Analysis}

Although no further data are being taken by the PHOBOS collaboration, analysis
work continues. One goal is to fully incorporate all results, including those
for smaller systems such as p+p and nucleus-nucleus collisions over an
extended range of centrality. Simultaneously, the analysis is expanding beyond
event-integrated single-particle distributions to the consideration of more
complicated observables such as fluctuations, correlations, and rare event
topologies. Results for elementary systems, to be used as a baseline
comparison for nucleus-nucleus data, have already been
published.\cite{ppCorrel}

\vspace{-8pt}
\section {Summary}

Analysis of the characteristics of particle production in nucleus-nucleus
collisions at RHIC energies have revealed a number of unexpectedly simple
dependencies. Observables considered range from the most basic such as total
multiplicity to the fairly complex such as elliptic flow. In many cases,
the dependencies on collision energy, centrality, pseudorapidity, and
transverse momentum factorize to a surprising degree. To paraphrase a comment
originally made about star formation in galaxies\cite{Noeske}, ``Particle
production in heavy ion collisions follows a quite simple pattern and simple
patterns often mean that there are only [a] few basic physical mechanisms at
work. \ldots We can now find out what these mechanisms are by measuring how
particle production behaves with energy, centrality, $\eta$, and p$_T$ and
compare that behavior to models". Extrapolation of these trends to LHC
energies suggest that interesting discoveries may well be made using only these
simply global observables.

\vspace{-8pt}
\section*{Acknowledgments}
This work was partially supported by U.S. DOE grants 
DE-AC02-98CH10886,
DE-FG02-93ER40802, 
DE-FG02-94ER40818,  
DE-FG02-94ER40865, 
DE-FG02-99ER41099, and
DE-AC02-06CH11357, 
U.\ S.\  NSF grants 9603486, 
0072204,            
and 0245011,        
Polish KBN grant 1-P03B-062-27(2004-2007),
NSC of Taiwan Contract NSC 89-2112-M-008-024, and
Hungarian OTKA grant (F 049823).

\end{document}